\documentclass[journal]{IEEEtran}

\usepackage{amsmath,amssymb,amsfonts,graphicx,mathtools,nccmath,amsthm,float}
\usepackage{multicol,tikz,cite,array,booktabs,url}
\usepackage{cite}
\usepackage{textcomp}
\usepackage{xcolor}
\usepackage{algorithm,comment}
\usepackage{algpseudocode,algorithmicx}
\usepackage{acronym}
\usepackage[inline]{enumitem}
\usepackage{subfigure}
\usepackage[bookmarks,colorlinks]{hyperref}

\begin{document}

\title{Smart Wireless Environments Enabled by RISs: \\Deployment Scenarios and Two Key Challenges}

\author{
    George C. Alexandropoulos\IEEEauthorrefmark{1}, Maurizio Crozzoli\IEEEauthorrefmark{2},  
    Dinh-Thuy Phan-Huy\IEEEauthorrefmark{7},
    Konstantinos D. Katsanos\IEEEauthorrefmark{1},\\
    Henk Wymeersch\IEEEauthorrefmark{5},
    Petar Popovski\IEEEauthorrefmark{8},
    Philippe Ratajczak\IEEEauthorrefmark{7},
    Yohann B\'{e}n\'{e}dic\IEEEauthorrefmark{7},
    Marie-Helene Hamon\IEEEauthorrefmark{7}, \\
    Sebastien Herraiz Gonzalez\IEEEauthorrefmark{7},
    Raffaele D'Errico\IEEEauthorrefmark{6}, and Emilio Calvanese Strinati\IEEEauthorrefmark{6}\\
    
    
    \IEEEauthorrefmark{1}National and Kapodistrian University of Athens, Greece, \IEEEauthorrefmark{2}TIM, Italy, \IEEEauthorrefmark{7}Orange Labs, France,\\
    \IEEEauthorrefmark{5}Chalmers University of Technology, Sweden,
    \IEEEauthorrefmark{6}CEA-Leti, France, \IEEEauthorrefmark{8}Aalborg University, Denmark
} 

\maketitle

\begin{abstract}
Reconfigurable Intelligent Surfaces (RISs) constitute the enabler for programmable propagation of electromagnetic signals, and are lately being considered as a candidate physical-layer technology for the demanding connectivity, reliability, localization, and sustainability requirements of next generation wireless communications networks. In this paper, we present various deployment scenarios for RIS-enabled smart wireless environments that have been recently designed by the ongoing EU H2020 RISE-6G project. The scenarios are taxonomized according to performance objectives, in particular, connectivity and reliability, localization and sensing, as well as sustainability and secrecy. We identify various deployment strategies and sketch the core architectural requirements in terms of RIS control and signaling, depending on the RIS hardware architectures and their respective capabilities. Furthermore, we introduce and discuss, via preliminary simulation results and reflectarray measurements, two key novel challenges with RIS-enabled smart wireless environments, namely, the \textit{area of influence} and the \textit{bandwidth of influence} of RISs, which corroborate the need for careful deployment and planning of this new technology.
\end{abstract}



\IEEEpeerreviewmaketitle

\section{Introduction} \label{sec:intro}
The potential of Reconfigurable Intelligent Surfaces (RISs) for programmable Electro-Magnetic (EM) wave propagation has recently motivated extensive academic and industrial interests, as an enabler for smart radio environments in the era of $6$-th Generation (6G) wireless networks \cite{RISE6G_COMMAG}. The RIS technology, which typically refers to artificial planar structures with almost passive electronic circuitry (i.e., without any power amplification), is envisioned to be jointly optimized with conventional wireless transceivers~\cite{huang2019reconfigurable} in order to significantly boost wireless communications in terms of coverage, spectral and energy efficiency, reliability, and security, while satisfying regulated EM Field (EMF) emissions.

The up-to-date RIS prototypes are composed of discrete sets of electromagnetically excited elements with tunable responses \cite{alexandg_2021}, and can be seen as either reflective or transmissive surfaces. Specifically, a reflective surface operates as an EM mirror, where an incident wave is reflected towards the desired direction (typically anomalous, in the sense that this direction can diverge from geometrical optics) with specific radiation and polarization characteristics. On the other hand, a transmissive RIS operates as a lens or a frequency selective surface, where the incident field is manipulated (by filtering, polarization change, beam splitting, etc.) and/or phase shifted, and re-radiated so as to control the refraction of impinging plane waves. It is noted that, when the RIS unit elements have both size and spacing lower than $1/10$th of the communication operating wavelength, RISs are also defined as metasurfaces \cite{Glybovski_2016}. Although RISs have great potential to implement advanced EM wave manipulations, mainly simple functionalities, such as electronic beam-steering and multi-beam scattering, have been demonstrated in the literature. Only very recently, RISs equipped with either minimal reception radio frequency chains \cite{alexandropoulos2021hybrid,HRIS,Alexandropoulos_2020a} or minimal power amplifiers \cite{Amplifying_RIS} have been introduced to enable sensing at the RIS side (which, apart from a standalone operation, can facilitate their network-wise orchestration) or reflection amplification (to confront the severe multiplicative pathloss), respectively.

The RIS-enabled programmability of information-bearing wireless signal propagation has recently gave birth to the concept of the \textit{Wireless Environment as a Service} \cite{rise6g}, which was introduced by the consortium of the EU H2020 RISE-6G project\footnote{See \url{https://RISE-6G.eu} for more information.}. This environment offers dynamic radio wave control for trading-off high throughput communications, energy efficiency, localization accuracy, and secrecy guarantees over eavesdroppers, while accommodating specific regulations on spectrum usage and restrained EMF emissions. However, the scenarios under which RISs will offer substantial performance improvement over conventional network architectures have not been fully identified. In this paper, capitalizing on latest studies within the RISE-6G project, we try to fill this gap and present various deployment scenarios for RIS-enabled smart wireless environments, while discussing their core architectural requirements in terms of RIS control and signaling, depending on the RIS hardware architectures and their respective capabilities. We also introduce two key novel challenges with RIS-enabled smart wireless environments, namely, the \textit{area of influence} and the \textit{bandwidth of influence} of RISs, which corroborate the need for careful deployment and planning of the technology.

\section{Scenarios for Connectivity and Reliability} \label{sec:conn_rel}
Conventional network scenarios impose communication performance to be achieved via uncontrolled wireless media, where the network is required to be tuned to one of the available service modes in an orthogonally isolated manner, offering non-focalized areas of harmonized and balanced performance. However, this might result in resource inefficiency and huge complexity. Conversely, the envisioned smart radio environment built upon RISs will enable the granting of highly localized quality of experience and specific service types to the end users \cite{rise6g}. In fact, such pioneering network paradigm aims at going one step beyond the classical $5$-th Generation (5G) use cases, by proposing performance-boosted areas as dynamically designed regions that can be highly localized, offering customized high-resolution manipulation of radio wave propagation to meet selected performance indicators. Although we do not cover the details in this paper, it is important to note that one of the central architectural elements of RIS-aided system is the control channel \cite{RISE6G_COMMAG}. The rate and latency of the control channel are crucial for timely and efficient configuration of RIS-empowered smart wireless environments, and should be in accord with the overall system requirements. 

For conventional system settings and strategies, based on a qualitative analysis of both localization feasibility (including possibly high-level identifiability considerations) and expected performance, the system engineer needs to determine where and how RISs can improve connectivity. In all scenarios described in the sequel, we consider that RISs are in reflecting operating mode. It is noted, however, that a recent line of extensive research \cite{alexandropoulos2021hybrid,HRIS,Alexandropoulos_2020a} focuses on RISs that can also receive signals to, for instance, perform in-band channel estimation or localization. This capability is expected to have an impact on the control channels for orchestrating smart wireless environments. Nevertheless, the deployments scenarios of sensing RIS will be similar to those of almost passive RISs, as long as the adoption of the former targets wave propagation control via radio-frequency sensing. 
Focusing on the downlink case as an example (the uplink case can be readily extended following the same approach), we taxonomize the RIS-empowered connectivity and reliability scenarios, as follows.
\begin{figure}[!t]
    \centering
    \subfigure[Connectivity and reliability boosted by a single RIS. ]{\includegraphics[width=0.45\textwidth]{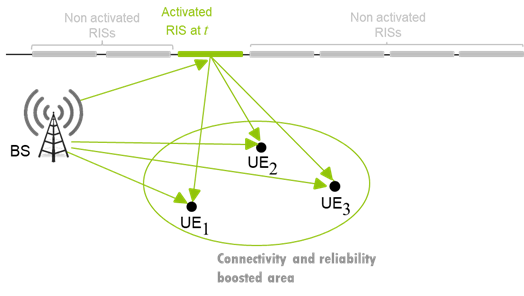}} \\
    \subfigure[RIS-empowered downlink communication of two BS-UE pairs, where each RIS can be controlled individually by each pair.]{\includegraphics[width=0.45\textwidth]{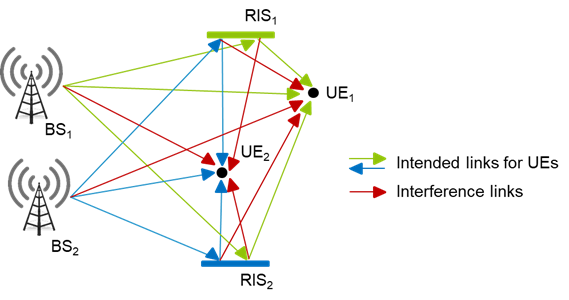}}
    \caption{Illustration of the connectivity and reliability scenarios \ref{sec:single_RIS} and \ref{sec:multi_RIS}.}
    \label{fig:scenarios_1_2}
\end{figure}

\subsubsection{Connectivity and reliability boosted by a single RIS} \label{sec:single_RIS}
In this scenario, there exist(s) direct link(s) between the multi-antenna Base Station (BS) and the single- or multi-antenna User Equipments (UEs). Connectivity is further boosted via a single RIS. The phase configuration/profile of the RIS can be optimized via a dedicated controller, who interacts with the BS, for desired connectivity and reliability levels.

\subsubsection{Connectivity and reliability boosted by individually controlled multiple RISs} \label{sec:multi_RIS}
In this scenario, multiple BSs aim at boosting connectivity and reliability with their respective UEs. The deployed RISs are assigned to BS-UE pairs, and each pair is capable to control and optimize the phase profile of its assigned individual RIS(s). In this case, the multiple RISs might result in multiplicative interference that needs to be carefully handled. 

\subsubsection{Connection reliability enabled by multiple RISs} \label{sec:multi_RIS2}
In this scenario, the direct link(s) between the multi-antenna BS and the single- or multi-antenna UE(s) is (are) blocked, and connectivity is enabled via a single or multiple RISs. Similar to scenario 1, the phase profile(s) of the RIS(s) can be optimized for desired reliability levels, requiring, however, forms of coordination among the different BSs (e.g., possibly similar to coordinated beamforming as in LTE-Advanced \cite{ltea}).

\subsubsection{Connectivity and reliability boosted by a single multi-tenant RIS} \label{sec:multi_ten_RIS}
For this scenario, we consider pairs of BS-UE(s) and a single RIS. The RIS is now considered as a shared resource, dynamically controlled by the infrastructure and commonly accessed by the BS-UE(s) pairs. The phase profile of the RIS can be commonly optimized by the BSs to serve their own UE(s) simultaneously. Alternatively, the control of the RIS may be time-shared among the BS-UE(s) pairs. Of course, the control channel envisioned by this scenario will have to be thoroughly investigated in future activities. A special case of this scenario is the one which considers a setup where the communication is enabled by multiple cellular BSs, each one serving a distinct set of UEs. When the UE(s) move across the cell boundaries of two or more BSs, they might change their serving BS(s) frequently (i.e., yielding frequent handovers). Shared RISs among BSs can be placed in the cell boundaries in order to dynamically extend the coverage of the serving BSs (i.e. reducing the number of handovers).

\subsubsection{Mobile edge computing as key-enabler in RIS-empowered scenarios} \label{sec:mec_RIS}
With the advent of beyond 5G networks, mobile communication systems are evolving from a pure communication framework to enablers of a plethora of new services, such as Internet of Things (IoT) and autonomous driving. These new services present very diverse requirements, and they generally involve massive data processing within low end-to-end delays. In this context, a key enabler is Mobile Edge Computing (or Multi-Access Edge Computing, namely MEC), whose aim is to move cloud functionalities (e.g., computing and storage resources) at the edge of the wireless network, to avoid the relatively long and highly variable delays necessary to reach centralized clouds. MEC-enabled networks allow UEs to offload computational tasks to nearby processing units or edge servers, typically placed close to the access points, to run the computation on the UEs’ behalf. However, moving towards millimeter-wave and THz communications, poor channel conditions due to mobility, dynamicity of the environment, and blocking events, might severely hinder the performance of MEC systems. In this context, a strong performance boost can be achieved with the advent of RISs, which enable programmability and adaptivity of the wireless propagation environment, by dynamically creating service boosted areas where energy efficiency, latency, and reliability can be traded to meet temporary and location-dependent requirements of MEC systems.

\begin{figure}[!t]
    \centering
    \subfigure[RIS-aided systems where connectivity is enabled by multiple RISs.]{\includegraphics[width=0.45\textwidth]{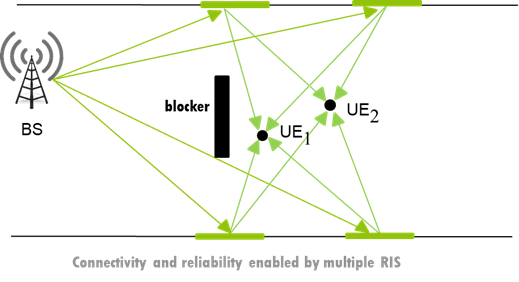}} \\
    \subfigure[A multi-tenancy scenario with two BS-UE pairs and a shared RIS that is optimized to simultaneously boost reliable communications.]{\includegraphics[width=0.45\textwidth]{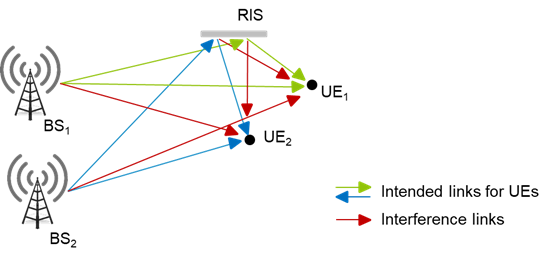}}
    \caption{Illustration of the connectivity and reliability scenarios \ref{sec:multi_RIS2} and \ref{sec:multi_ten_RIS}.}
    \label{fig:scenarios_3_4}
\end{figure}
\section{Scenarios for Localization and Sensing} \label{sec:local_sens}
RIS technology is expected to enable advanced sensing and localization techniques for environment mapping, motion detection, opportunistic channel sounding, and passive radar capabilities applied to industrial (e.g. smart factory), high user-density (e.g. train stations), and indoor (e.g. augmented/mixed reality) environments. 
%
%
The expected benefits can take various forms, depending on the RIS operating mode (i.e., reflect, refract, transmit, relay, etc.), and they can be classified in terms of \textit{i}) {enabled localization} (i.e.,  making localization feasible where  the conventional system  \cite{delPeral_Rosado_2018,Keating_2019} fails;
 \textit{ii}) {boosted localization} (i.e., improving the localization performance); and \textit{iii})
{low-profile localization} (i.e.,  localization 
requiring much lower resources in comparison with the conventional system). 

In the following sections, ten generic scenarios where RISs provide performance benefits for localization are presented 
(refer to the technical literature for a more comprehensive overview \cite{Keykhosravi_2021,Abu_Shaban_2021,Keykhosravi_2021a}).

\subsubsection{Unambiguous localization under favourable problem geometry with a minimal number of BSs}
A single RIS enables localization \cite{Keykhosravi_2021}, while 
multi-RIS deployment setting can also contribute to boost performance, by a combination of time and angle measurements. 

\subsubsection{Non Line-of-Sight (LOS) mitigation for better service coverage and continuity in far-field conditions}
Whenever the minimum number of BSs (anchors) in visibility is not fulfilled, the UE location can be estimated via narrowband signals received from two RISs. 

\begin{figure*}[!t]
\centering
  \includegraphics[width=15cm, height=8cm]{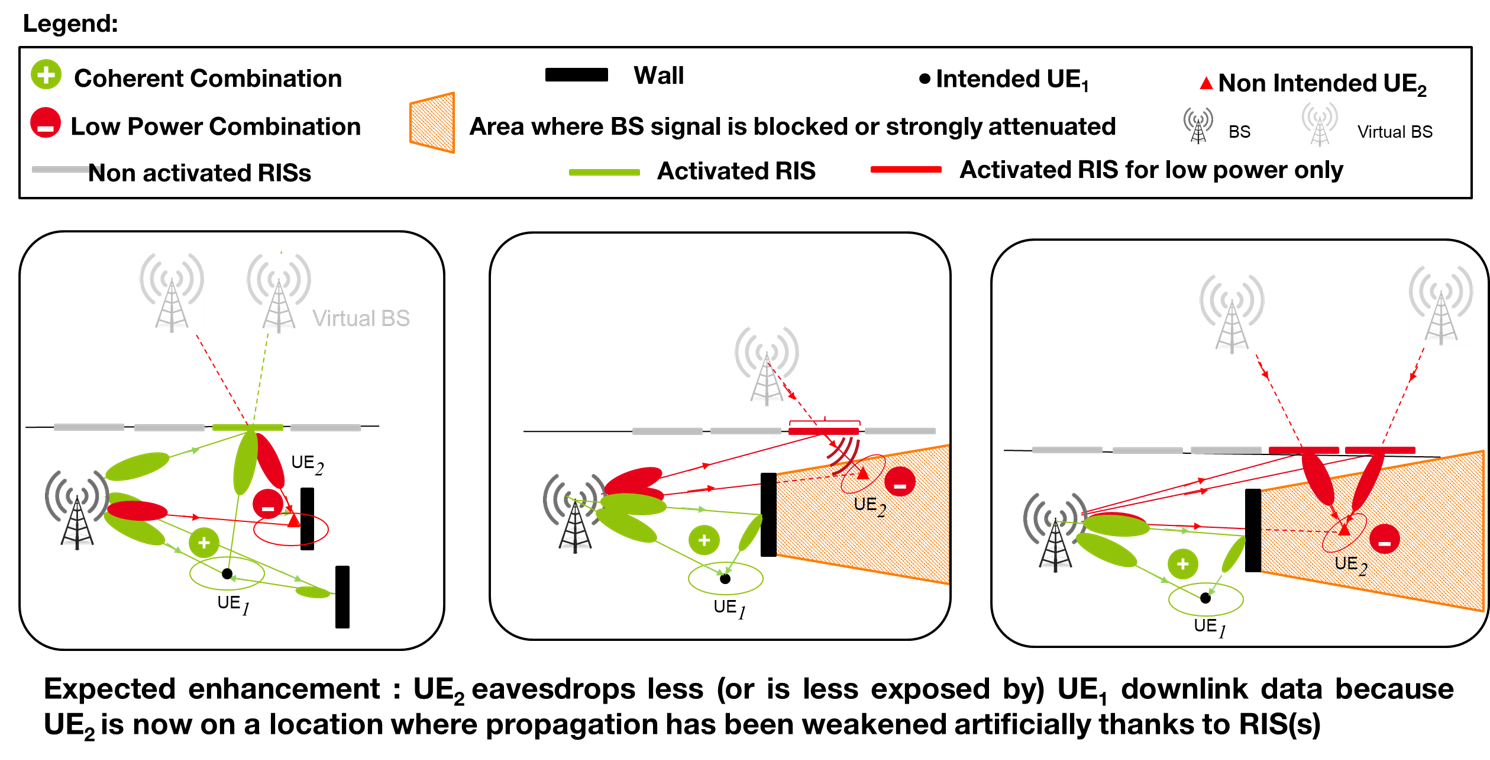}
  \caption{The three core scenarios for RIS-empowered sustainability and secrecy discussed in Section~\ref{sec:sust_secur}.}
  \label{fig:scenarios_4_3}
\end{figure*}
\subsubsection{Non LOS mitigation for better service coverage and continuity in near-field conditions}
Extending the previous scenario, the UE location can be estimated via the signal received from one RIS even without any BS in visibility, 
when the user is in the near-field of the RIS. This allows to exploit signal wavefront curvature for direct positioning \cite{Abu_Shaban_2021,Rahal_2021}. 

\subsubsection{On-demand multi-user and multi-accuracy service provision}
The deployment (and the selective control) of multiple RISs makes possible \textit{i}) the (on-demand) provision of various classes of localization services to different UEs sharing the same physical environment, depending on the needs they express locally/temporarily, while \textit{ii}) spatially controlling both the localization accuracy and the geometric dilution of precision in different dimensions (i.e., both the sizes and orientations of the location uncertainty ellipses) \cite{Wymeersch_2020}. 


\subsubsection{Opportunistic detection/sensing of passive objects through multi-link radio activity monitoring}
Similar to standard range-Doppler analysis \cite{Li_2019}, 
this is possible by monitoring the time evolution of multipath profiles over a communication link between the BS and one or several UEs, harnessing the dynamic and selective control of RISs \cite{Lu_2021,Jiang_2021b,Buzzi_2021b}. 
In highly reverberant environments, 
multiple RISs can create 
configurational diversity through wavefront shaping, even with single-antenna single-frequency measurements \cite{alexandg_2021}.

\subsubsection{RIS-assisted search-and-rescue operations in emergency scenarios}
RISs may support and overcome the shadowing effect induced by rubble in such scenarios by building ad-hoc controllable propagation conditions for the cellular signals employed in the measurement process \cite{Albanese_2021}. 
In addition, lightweight and low-complexity RISs 
mounted on drones can be used to bring connectivity capabilities to hard-to-reach locations, supporting first responders \cite{Mursia_2021}. 

\subsubsection{Localization without BSs using a single or multiple RISs}
In this scenario, wideband transmissions to a single passive RIS \cite{zerobs} or angle estimations obtained at multiple RISs \cite{locrxris}, having the architecture of \cite{Alexandropoulos_2020a}, can be combined to produce the estimation of UE(s) location(s).

\subsubsection{RIS-aided radio environment mapping for fingerprinting localization}
RISs equipped with minimal reception circuitry for sensing \cite{Alexandropoulos_2020a} enable the cartography of the electromagnetic power spatial density in a specific area of interest, 
accentuating the location-dependent features of the RIS-enhanced radio signatures stored in the database. 

\subsubsection{RIS lens}
In this scenario, the RIS is placed in front of a single-antenna transmitter. The user position is estimated at the UE side via the narrow-band received signal, exploiting wavefront curvature \cite{Abu_Shaban_2021}.

\subsubsection{Radar localization/detection of passive target(s) with hybrid RIS}
Using the architecture of \cite{HRIS}, a radar is assisted by multiple hybrid RISs in order to localize/detect both static or moving target(s), extending scenario 5. 
Hybrid RISs 
simultaneously receive and reflect, 
giving the system the capability to localize UEs as well as the radar itself.


\section{Scenarios for Sustainability and Secrecy} \label{sec:sust_secur}
RIS-empowered smart radio environments are expected to improve the sustainability and security of 6G networks by focusing the energy towards the target UE, and reducing the amount of unnecessary radiations in non-intended directions in general, or in the directions of non-intended UEs (e.g., exposed UEs or eavesdroppers). More precisely, the following metrics can be improved: \textit{i}) the Energy Efficiency (EE) \cite{huang2019reconfigurable} (for instance, this metric can be defined as the attained data rate at the target UE divided by the BS transmit power); \textit{ii}) the EMF utility (for instance, this metric can be defined as the attained data rate at the target UE divided by the largest EMF exposure at another UE); and \textit{iii}) the Secrecy Spectral Efficiency (SSE) \cite{PLS_Kostas} (for instance, this metric can be defined as the attained data rate at the target UE minus the data rate that is intercepted by the eavesdropper). A first already identified use case for such improvements \cite{rise6g} is a ``train station,'' where an orchestrated network of RISs can be optimally configured to maximize the aforementioned metrics, in small areas close to the surfaces. These areas can be advertised to customers as ``areas with high EE,'' ``areas with low EMF exposure,'' or ``areas with increased secrecy.''

\begin{figure*}[!t]
\centering
  \includegraphics[scale=0.27]{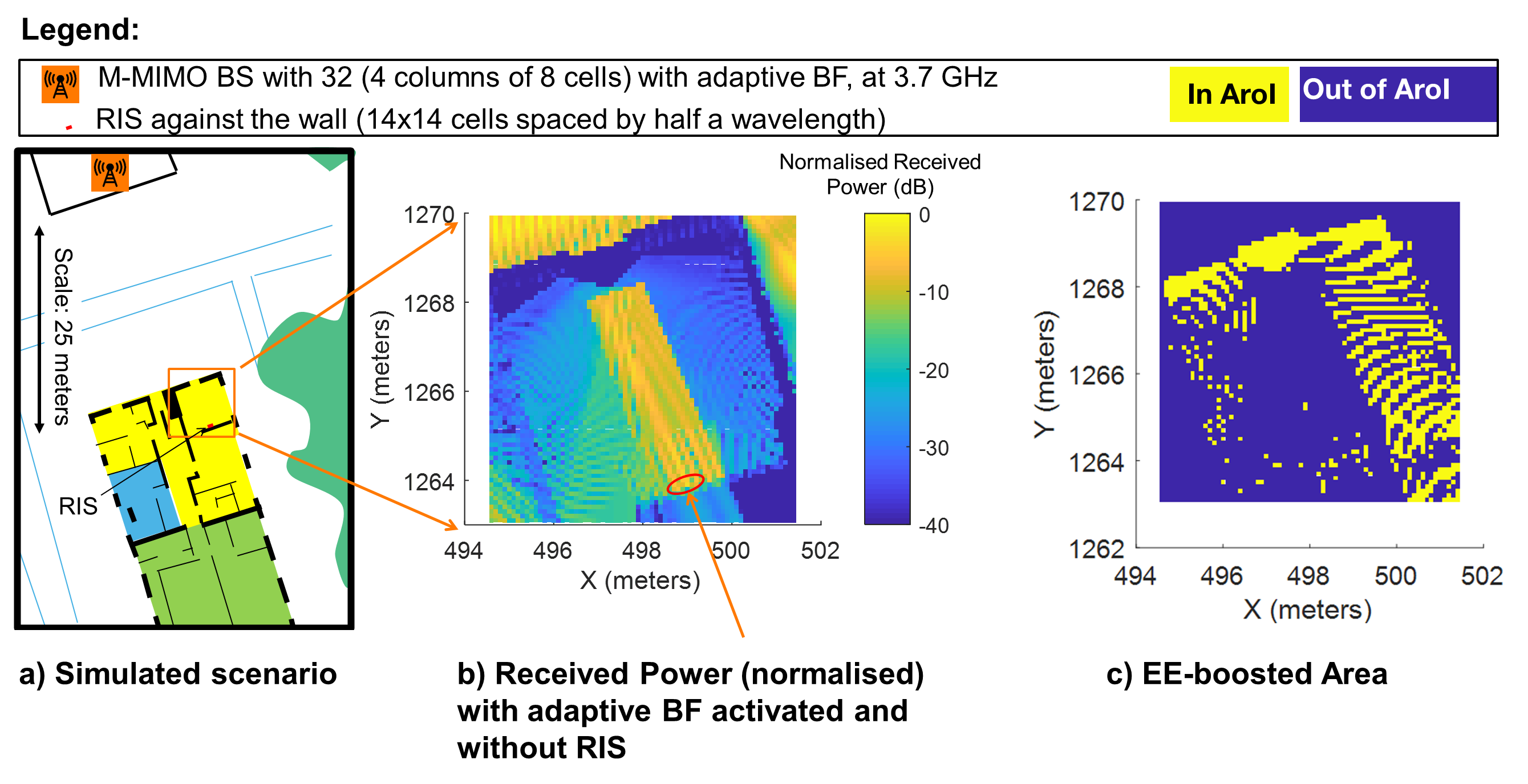}
  \caption{Simulation example of an EE-boosted area in an outdoor-to-indoor environment, where the received power at the target UE has been boosted by at least $3$ dB in the Area of Influence (AroI) of the RIS.}\vspace{-0.1cm}
  \label{fig:AroI}
\end{figure*}
\begin{figure*}[!t]
  \centering
  \subfigure[SSE without an RIS.]{\includegraphics[scale=0.5]{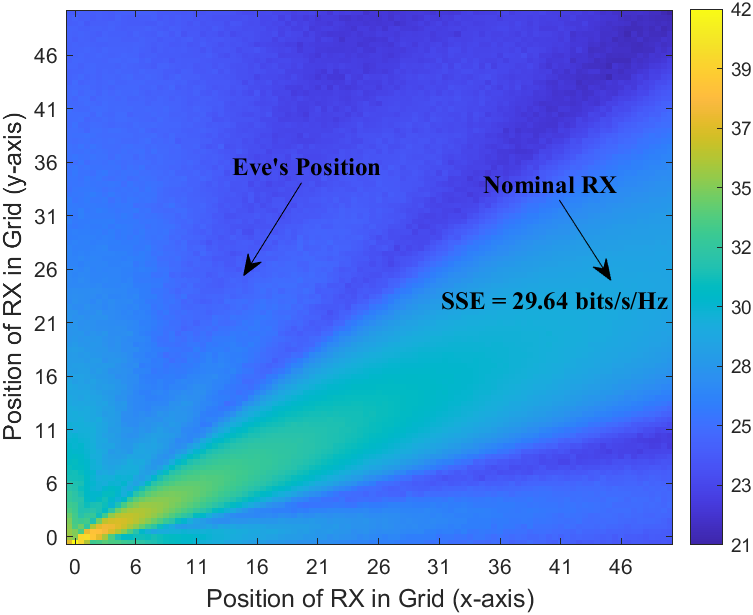}}\quad
  \subfigure[SSE with an RIS.]{\includegraphics[scale=0.5]{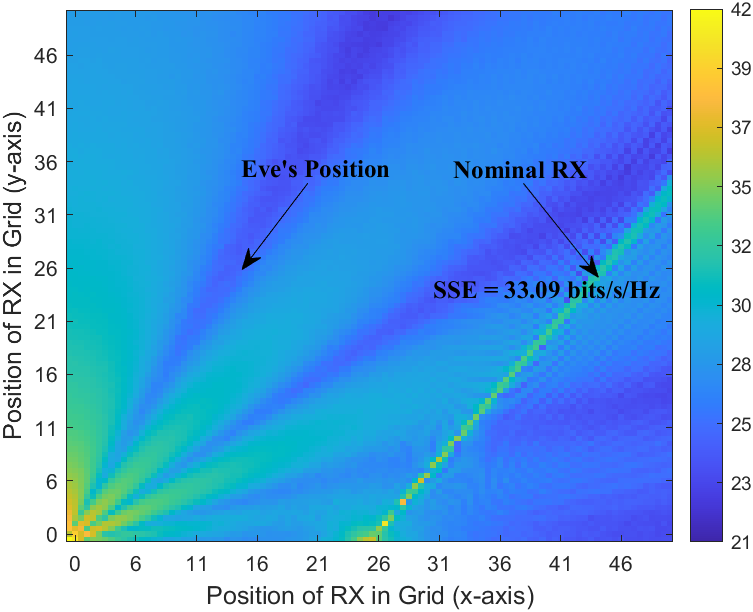}}
  \caption{The Secrecy Spectral Efficiency (SSE) performance with joint BF and RIS configuration optimization, and the resulting AroI of the RIS.}\vspace{-0.1cm}
  \label{fig:SSE_SpatialFoc}
\end{figure*}

In contrast to conventional scenarios without deploying RISs, we illustrate in Fig.~\ref{fig:scenarios_4_3} examples of single-BS scenarios with RIS(s), where downlink transmit BeamForming (BF) is used to boost the received power at the target intended UE and to reduce the received power at the non-intended UE; this can be achieved by exploiting the artificial shaping of the propagation channels thanks to RIS(s). When the non-intended UE is an exposed UE, the obtained link is a low EMF link, whereas when the non-intended UE is an eavesdropper, the obtained link is a secured link. In the figure, the advantages brought by RIS(s) to reduce the received power at the non-intended UE (whether it is an eavesdropper or an exposed UE) compared to the received power at the target UE are illustrated for three types of propagation scenarios. 
\begin{enumerate}
    \item \textit{The BS-to-intended UE link is in LOS}: In this case (left figure), the RIS artificially adds propagation paths to the channel, which coherently combine with other ``natural'' paths to boost the received power at the target UE, and non-coherently combine with other ``natural'' paths to reduce the received power at the non-intended UE.
    \item \textit{The BS-to-intended UE link is blocked by an obstacle and the intended UE is in the near-field of an RIS}: In this case (middle figure), the RIS artificially adds a propagation path to the existing ``natural'' paths to reduce the received power at the non-intended UE;
    \item \textit{The BS to intended UE link is blocked by an obstacle and the intended UE is in the far-field of an RIS}: This case (right figure) is similar to the previous case except that several RISs may become useful to reduce the energy at the non-intended UE.
\end{enumerate}
Note that, in the absence of a non-intended UE, the proposed system in Fig.~\ref{fig:scenarios_4_3} only boosts the received power at the target UE, and therefore, boosts the EE metric. Note also that similar scenarios for uplink can be derived with receive BF instead of transmit BF. Regarding EE, the main difference will be that the EE will be improved at the UE side. Regarding EMF utility, the main difference will be that the target UE and the exposed UE will be the same person, instead of being distinct people. In this case, the RISs will help reducing the self-exposure of a UE transmitting data in the uplink with its own smartphone. Furthermore, the previously described scenarios can be generalized to multi-BS scenarios, where several synchronized and coordinated BSs perform joint beamforming.

\begin{figure*}[!t]
\centering
  \includegraphics[scale=0.48]{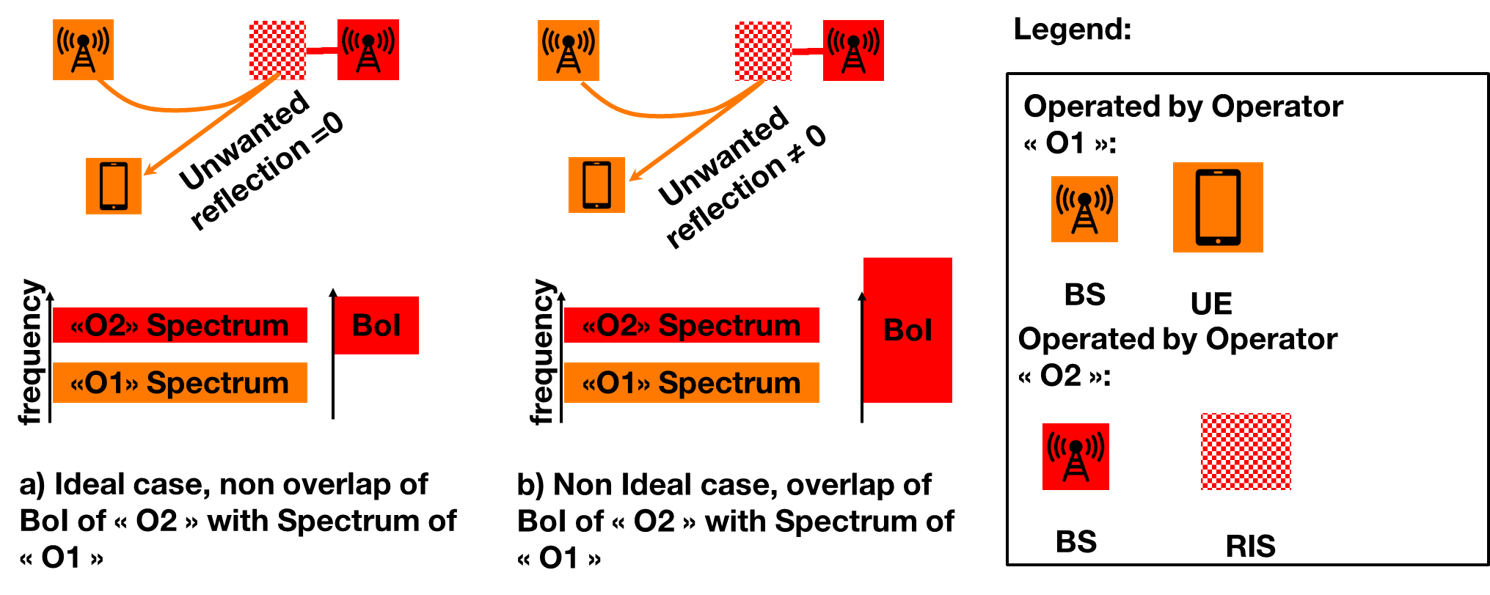}
  \caption{The RIS Bandwidth of Influence (BoI) and the resulting undesired reflections.}
  \label{fig:BoI}\vspace{-0.16cm}
\end{figure*}

\section{The AroI and BoI Challenges} \label{sec:aroi_boi}
We have simulated an outdoor-to-indoor wireless signal propagation scenario using Orange's ray-tracing tool, as depicted in Fig.~\ref{fig:AroI}a). In this scenario, there exists an outdoor Massive Multiple-Input Multiple-Output (M-MIMO) BS which applies Maximal Ratio Transmission (MRT) to serve UEs placed inside the highlighted room. Figure~ \ref{fig:AroI}b) illustrates the effective coverage attained in the room, in terms of received power, when BF gain is taken into account and no RIS is deployed. We then compute the gain in received power, and therefore the gain in EE, when using an RIS installed against the most illuminated wall (by the outdoor BS); the phase configuration of this RIS has been jointly optimized with the MRT BF \cite{NA2021}. In Fig.~\ref{fig:AroI}c), the area where this gain in received power (equivalently, this boosts EE) exceeds $3$ dB is colored in yellow. This area can be seen as a kind of AroI of the RIS on the coverage of the BS. We have actually observed that the shape of this area is far from being easily predictable. It accounts for the complex propagation between the BS and the room, between the RIS and locations inside the room, as well as the joint optimization of BF and RIS configuration. Therefore, RIS deployment and planning constitute a key challenge for RIS-empowered smart wireless environments, and may require new planning methods and tools. Finally, in Fig.~\ref{fig:SSE_SpatialFoc}, we illustrate the SSE performance boosting in a strong LOS environment offered by RIS optimization using \cite{PLS_Kostas}, and the resulting AroI of the RIS.

In \cite{PR2013_b}, measurements have shown that a reflectarray prototype, optimized for $5.2$ GHz, does reflect incident signals in a wide bandwidth of $1$ GHz. This bandwidth can be seen as the BoI of the reflectarray over any incident signal. To this end, Fig.~ \ref{fig:BoI} illustrates the scenario where two operators O1 and O2, have distinct spectrum allocations, and O2 deploys an RIS. With an ideal RIS, the BoI of the RIS perfectly matches O2's spectrum, and therefore, the signals coming from O1 and hitting the RIS won't be subject to unwanted re-radiations or reflections. On the contrary, if the BoI overlaps with O1's spectrum, the signals from O1 and hitting the RIS will undergo undesired re-radiations. This may potentially create unexpected perturbations in the coverage of O1. Therefore, coexistence between operators in the presence of RIS(s) may be a challenge and may require new technical solutions, which may have an impact to spectrum regulation.

\section{Conclusion} \label{sec:concl}
In this paper, we presented various deployment scenarios for RIS-enabled smart wireless environments, as they have been identified by the ongoing EU H2020 RISE-6G project. In particular, the scenarios have been categorized according to the expected achievable performance, and their goal was to lead in substantial performance improvement compared to conventional network architectures. However, similar to any promising new technology, the increased potential of RISs is accompanied with certain challenges. Two novel ones are the AroI and BoI, which have been introduced and discussed via preliminary simulations results and measurements.

\section{Acknowledgement}
This work has been supported by the EU H2020 RISE-6G project under grant number 101017011.

\bibliographystyle{IEEEbib}
\bibliography{IEEEabrv, EuCNC_Refs_v1}

\end{document}